\documentclass[10pt,a4paper]{article}

\usepackage{moreverb,xspace,bm}
\usepackage[left=3cm,right=2cm,top=3cm,bottom=3cm]{geometry}

\usepackage[affil-it]{authblk}
 
\usepackage{natbib}
\usepackage{amsmath,amssymb,amsthm,graphicx,natbib,bbm,multirow,rotating}
\usepackage{parskip}
\usepackage{float,subfig}
\usepackage{multicol}

\newtheorem{theorem}{Theorem}[section]

\newtheorem{lemma}[theorem]{Lemma}

\newtheorem{condition}[theorem]{Condition}
\newtheorem{property}[theorem]{Property}
\newtheorem{definition}[theorem]{Definition}

\newcommand{\cip}{\,\perp\!\!\!\perp}
\newcommand{\nip}{\,\,\not\!\perp\!\!\!\perp}
\newcommand{\as}[1]{\mbox{\rm a.s. [$#1$]}}
\newcommand{\cS}{{\cal S}}

\newcommand{\mE}{\mathbb{E}}
\newcommand{\mP}{\mathbb{P}}

\newcommand{\indo}[2]{\mbox{$#1 \,\cip\, #2$}}
\newcommand{\ind}[3]{\mbox{$#1 \, \cip\, #2 \mid #3$}}
\newcommand{\nind}[3]{\mbox{$#1 \, \nip\, #2 \mid #3$}}

\renewcommand{\eqref}[1]{\mbox{(\ref{eq:#1})}}

\newcommand{\defref}[1]{\mbox{Definition~\ref{def:#1}}}

\newcommand{\thmref}[1]{\mbox{Theorem~\ref{thm:#1}}}
\newcommand{\propref}[1]{\mbox{Property~\ref{prop:#1}}}
\newcommand{\condref}[1]{\mbox{Condition~\ref{cond:#1}}}

\makeatletter
\newcommand{\xRightarrow}[2][]{\ext@arrow 0359\Rightarrowfill@{#1}{#2}}
\makeatother

\theoremstyle{plain}

\title{Regression Discontinuity Designs: A Decision Theoretic Approach} 

\author[1]{Panayiota Constantinou}

\author[2]{Aidan G.~O'Keeffe%
\thanks{E-mail address: \texttt{a.o'keeffe@ucl.ac.uk}; Corresponding author}}

\affil[1]{\small School of Mathematics, University of Bristol, BS8 1TH, UK.}
\affil[2]{\small Department of Statistical Science, University College London, WC1E 6BT, UK.}

\begin{document}

\maketitle

\begin{abstract}
The regression discontinuity design (RDD) is a quasi-experimental design that can be used to identify and estimate the causal effect of a treatment using observational data. In an RDD, a pre-specified rule is used for treatment assignment, whereby a subject is assigned to the treatment (control) group whenever their observed value of a specific continuous variable is greater than or equal to (is less than) a fixed threshold. Sharp RDDs occur when guidelines are strictly adhered to and fuzzy RDDs occur when the guidelines or not strictly adhered to. In this paper, we take a rigorous decision theoretic approach to formally study causal effect identification and estimation in both sharp and fuzzy RDDs. We use the language and calculus of conditional independence to express and explore in a clear and precise manner the conditions implied by each RDD and investigate additional assumptions under which the identification of the average causal effect at the threshold can be achieved. We apply the methodology in an example concerning the relationship between statins (a class of cholesterol-lowering drugs) and low density lipoprotein (LDL) cholesterol using a real set of primary care data.   

\end{abstract}


\section{Introduction}
\label{sec:1}

Regression discontinuity designs (RDDs) were first developed during the 1960s with the aim of estimating the causal effect of an intervention using observational data (\cite{thi:60}). Such designs occur where a decision to apply a treatment is linked to some continuous `assignment variable' through a decision rule. Here, a `treatment' refers to any action that is taken with the aim of affecting some response in an outcome variable of interest. For example, in an educational context, the `treatment' might be the assignment of students to a particular school based on an admissions test result (the assignment variable), where students are given a place at the school based on their attainment of a test `pass mark' (the decision rule). In a medical context the `treatment' might be the prescription of a drug based on the value of a subject's diagnostic test result (the assignment variable). Typically, a decision rule takes the form of a `treatment threshold' whereby a subject receives the treatment if the subject's assignment variable value lies at or above the treatment threshold and the subject does not receive the treatment if the subject's assignment variable lies below the treatment threshold, at a pre-specified point in time. 

Assuming that subjects with similar assignment variable values are exchangeable and that the decision rule is identical for all subjects, a comparison of outcome variable values between subjects whose assignment variables lie `just above' the threshold and `just below' the threshold might be considered appropriate for the calculation of a causal effect of the treatment on the outcome of interest. For example, in a medical context, statin therapy - that is, prescription of lipid-modifying drugs (the treatment) may be prescribed to patients only if their 10-year risk of developing cardiovascular disease (the assignment variable) lies above a pre-specified threshold of 20\% (the treatment threshold). Patients whose risk scores lie close to 20\% are perhaps likely to be similar with respect to their age, lifestyle and other variables implying that we might have a population of exchangeable subjects to whom an external decision rule is applied to determine allocation to statins. 

RDDs have been applied in many areas of economics and social science (see, for example, \citet{ber:99}, \citet{van:02}, \citet{van:08}, \citet{lal:08}). RDDs have not been used extensively in medicine and epidemiology, although their use and development is increasing in this field (\citet{lin:06}, \citet{alm:11}, \citet{lin:12}, \citet{bor:14}, \citet{van:14}, \citet{oke:14}). The increasing availability of large observational healthcare databases suggests that the RDD could be considered more widely and developed for use in biostatistics and epidemiology.    

Decision theory has been developed and used as an approach to causal inference (\citet{rub:78}, \citet{hec:95}, \citet{daw:00}). By definition, an RDD includes a decision rule and, as such, the link between the decision theoretic approach to causal inference and the RDD appears natural. A decision theoretic approach typically involves a set of assumptions made about the probabilistic behaviour of variables of interest across a variety of regimes. Different regimes might refer to, for example, different time points, locations or contexts. In our context, we consider a regime to be either `interventional', when it results from an external intervention (for example, by the statistician) which forces the treatment variable to take on a particular value, or `observational', when it results from an observed setting where the treatment was determined by a number of known or unknown factors (e.g.~doctor or patient preference for treatment, governmental guidelines etc.). RDDs represent examples of `observational regimes' and are commonly classified as either `sharp' or `fuzzy'. In a sharp RDD, subjects are assigned the treatment strictly according to the treatment threshold. In other words, at the point where the decision to assign a treatment is taken, all subjects who are given the treatment will exhibit an assignment variable value at or above the treatment threshold and, conversely, all subjects who do not receive the treatment will exhibit an assignment variable value below the treatment threshold. In a fuzzy RDD, this might not be the case; there may be some subjects whose receipt (or not) of the treatment is contrary to what is indicated by the decision rule. Figure \ref{plot1} shows plots of example fuzzy and sharp RDDs, for a continuous assignment variable $x \in \mathcal{X}\subseteq\mathbb{R}$ such that a subject receives the treatment if $x \geq x_{0}$, where $x_{0} \in \mathcal{X}$ defines the treatment threshold level. In these example plots, we assume a continuous outcome variable of interest, denoted $y\in\mathcal{Y}\subseteq\mathbb{R}$.

In this work, we define and formulate a decision-theoretic approach to the RDD which, to our knowledge, has not been attempted. We use ideas of  conditional independence (\citet{daw:79}, \citet{daw:80}) to describe assumptions concerning the variables defined in a RDD and how these assumptions relate to considered regimes. We derive appropriate causal effect estimators for sharp and fuzzy RDDs and present examples of treatment effect evaluation using a real example concerning the initiation of statin therapy for the prevention of cardiovascular disease in UK primary care.

\section{Notation and Terminology within a Decision Theoretic framework}
\label{sec:2}
In this section, we define the notation and terminology that will be used within the Decision Theoretic (DT) framework of statistical causality. The DT framework was introduced by \citet{daw:00} and has been used to address a variety of problems in causality \citep{gen:07,dawdid:10, dawcon:14}. In this framework, we acknowledge that the distributions of stochastic variables of interest are different, in principle, between different regimes (observational or interventional). We explore relationships that may be assumed between the differing distributions using the language and calculus of conditional independence \citep{daw:79, daw:80, con:14}. The example that will be used throughout the paper is the one described previously, where statins may be prescribed to patients whose risk of developing cardiovascular disease (CVD) within 10 years exceeds $20 \%$. Our objective is to make causal inference about the effect of the treatment for those individuals who have measurements around the threshold and thus inform policy makers about the effectiveness of statins on lowering LDL cholesterol when prescribed according to the guidelines. We note that everything we describe using this specific example, may be generalised to a variety of contexts.

First, we consider the treatment variable $T$ (prescription of statins). For simplicity, we consider $T$ to be binary, taking the value 0 when the treatment is not administered and value 1 when the treatment is administered. We consider $X$ to be a continuous assignment variable (10-year CVD risk score) using which the treatment guidelines are applied. In addition, we consider a threshold $x_0$, where, when $X \geq x_0$ the guidelines recommend treatment and when $X < x_0$ the guidelines recommend no treatment. In our example, $x_0=20 \%$, as recommended by United Kingdom's National Institute for Health and Care Excellence (NICE) guidelines (\citet{NICE}). Associated with these measurements is the binary threshold indicator $Z$ which denotes the choice for treatment as imposed by the guidelines. So $Z=0$, when $X < x_0$ and $Z=1$, when $X \geq x_0$. We consider the response variable of interest (LDL cholesterol level) and denote this by $Y$. We aim to identify the causal effect of $T$ on $Y$ and estimate this for patients around the threshold. Further to the above variables, we consider confounding variables; in other words, variables that can influence the treatment choice and at the same time affect the response variable $Y$. We denote these variables by $C$ and note that a subset of $C$ may be taken into account for determining $X$. In our example, some variables such as age, gender, smoking status, cholesterol levels,  etc., are those using which the 10-year risk score is calculated but some variables, such as a doctor's preference for treatment, patient's preference for treatment, allergies, reactions to previous treatments, etc., might be additional to those encompassed in $X$ (i.e., $X$ is a function of $C$). In the sharp design none of these variables (except through $X$) can affect the treatment that will be administered. However, in the fuzzy design some of these variables might outweigh the guidelines, especially if the doctor believes that the guidelines did not put appropriate weight on specific measurements considered important and that the patient would benefit from a different treatment than that which the guidelines recommend. 

Further to the stochastic variables, we introduce a non-stochastic variable, $\Sigma$, which will indicate the operating regime. Interpreting as \textit{causal} the \textit{effect of an intervention}, we want to compare the distribution of the response variable $Y$, around the threshold $x_{0}$, between the two interventional regimes; the regime where we intervene to give treatment with the regime where we intervene not to give treatment. These two interventional regimes are considered otherwise identical. However, the regimes from which we obtain data (sharp or fuzzy RDD) do not necessarily represent interventional settings but only provide a (usually incomplete) record of observations as they have been generated by Nature and collected by the GP/hospital. Thus, in order to make inference for the interventional regimes using information from the observational regime, we need to introduce and justify assumptions that connect the probabilistic behaviour of the random variables of interest across the differing regimes. Here, we emphasise that the regime indicator $\Sigma$ has no uncertainty associated with it, and serves only as a parameter to index the regimes. More precisely, 
\begin{equation*}
\Sigma = \left\{
   \begin{array}{rl}
   0, & \text{for the interventional regime under control treatment,}\\
   1, & \text{for the interventional regime under active treatment,} \\
   s, & \text{for the observational regime under a sharp RDD,} \\
   f, & \text{for the observational regime under a fuzzy RDD.} \\
   \end{array} \right.
\end{equation*}
We emphasise that any probabilistic statement concerning the stochastic variables should be made, explicitly or implicitly, conditional on the value of the regime. Thus, we will write $\mathbb{P}_{0}(Y)$ to denote the distribution of $Y$ under the interventional regime $\Sigma=0$, $\mathbb{E}_{s}(Y \mid X=x_0)$ to denote the conditional distribution of $Y$ given $X=x_0$, under the observational sharp RDD $\Sigma=s$, and use similar notation throughout the paper. Note that when an observational regime is in operation, the distribution of all the random variables involved, and in particular the treatment variable $T$, will be determined by Nature. On the contrary, when an interventional regime is in operation, the treatment $T$ will take the same value as the one of the regime with probability 1, i.e., $\mathbb{P}_{0}(T=0)=1$ and $\mathbb{P}_{1}(T=1)=1$. Hence, in the interventional regimes, $T$ is a degenerate random variable with all probability on one value, the same as the value of $\Sigma$. 

Our principal aim is to measure the impact of the treatment intervention on $Y$, at the threshold $x_0$. As such, the quantity we seek to identify and estimate is the {\em Average Causal Effect at the threshold $x_0$}, denoted by $ACE_{x_0}$, where

\begin{align}
\label{def:ACE}
ACE_{x_{0}} &:=\mathbb{E}_1(Y \mid X=x_{0})-\mathbb{E}_0(Y \mid X=x_{0}).
\end{align}

The $ACE_{x_0}$ is the difference in expectation of Y, between the two interventional regimes $\Sigma=0$ and $\Sigma=1$, at the threshold $x_0$. Thus, a non-zero value admits a causal interpretation. In the following sections, we will explore assumptions, expressed in the language of conditional independence, that allow the identification of the $ACE_{x_0}$ from the corresponding observational regime (sharp or fuzzy RDD). 

\section{Conditional independence}
\label{sec:3}
Conditional independence in a form that encompasses simultaneously stochastic and non-stochastic variables (extended conditional independence) was first introduced by \citet{daw:79, daw:80} and the required calculus, known as the axioms of conditional independence, has been established under conditions by \citet{con:14}. In this section, we give definitions for extended conditional independence as they will be used in the paper and state the properties that will be needed in the following section. Where theorems and lemmas are given, corresponding proofs are provided in the appendix for this paper.   

We denote by $\mathcal{S}$ the set of contemplated regimes, where $\mathcal{S}\subseteq\{0,1,s,f\}$. Also, $X_{1}, X_{2}, \ldots$ denote stochastic variables and $\Sigma$ denotes the non-stochastic regime indicator that takes values $\sigma$ in the set $\mathcal{S}$. To specify explicitly the underlying probability measure considered in regime $\sigma$, we write $\mathbb{P}_{\sigma}$.
~\\
\begin{definition}
\label{def:nstoc1}
We say that $X_{1}$ is (conditionally) independent of $X_{2}$ given $(X_{3},\Sigma)$ and write \ind {X_{1}}{X_{2}}{(X_{3},\Sigma)} if, for any $\sigma \in \cS$ and any (bounded, real and measurable) function $h(X_1)$,
  \begin{equation}
  \label{eq:nstoc1}
    \mE_{\sigma}[h(X_{1}) \mid X_{2},X_{3}]=\mE_{\sigma}[h(X_{1}) \mid X_{3}] \quad \as{\mP_{\sigma}}.
  \end{equation} 
\end{definition}

In other words, \ind {X_{1}}{X_{2}}{(X_{3},\Sigma)} if, separately in each regime $\sigma \in \cS$, the distribution of $X_{1}$ given $(X_{2},X_{3})$ depends, in fact, only on $X_{3}$. So under each regime, upon information on $X_{3}$, further information on $X_{2}$ becomes redundant for making probabilistic inference about $X_{1}$. If such a statement is not true, we will write instead \nind{X_{1}}{X_{2}}{(X_{3},\Sigma)}.
 ~\\
\begin{definition}
\label{def:nstoc2}
We say that \emph{$X_{1}$ is (conditionally) independent of $\Sigma$ given $X_{2}$}, and write \ind {X_{1}} {\Sigma} {X_{2}}, if for any (bounded, real and measurable) function $h(X_{1})$, there exists a function $w(Z)$ such that, for all $\sigma \in {\cS}$,
  \begin{equation}
    \label{eq:nstoc2}
    \mE_{\sigma}[h(X_{1})\mid X_{2}]=w(X_{2}) \quad \as{\mP_{\sigma}}.
  \end{equation}
\end{definition}

Interpreting the above definition, $\ind {X_{1}} {\Sigma} {X_{2}}$ requires that, there exists a function $w(X_2)$, which does not depend on the regime, and can serve as a version of the conditional expectation $\mathbb{E}[h(X_{1})\mid X_{2}]$ simultaneously in all regimes. So \eqref{nstoc2} provides a link between the (conditional) distribution of $X_{1} \mid X_{2}$ in all $s \in \cS$. Intuitively, \ind {X_{1}} {\Sigma} {X_{2}} reflects the understanding that, upon information on $X_{2}$, further information on the operating regime becomes irrelevant for making probabilistic inference about $X_{1}$. Similarly to the above, if such a statement is not true, we will write instead  \nind{X_{1}}{\Sigma}{X_{2}}.

Starting with a set of assumptions (which we believe represent the problem under study), we want to explore if they lead to identification of the $ACE_{x_0}$ from observational data. Identification can follow directly (by definition of such assumptions) or indirectly (by further properties which can be induced by such assumptions). In order to explore such paths, we use the axioms of conditional independence. In \thmref{axioms}, we write  $W \preceq Y$ to mean that $W$ is a function of $Y$.
~\\
\begin{theorem}[Axioms of conditional independence]
\label{thm:axioms}
Let $X,Y,Z,W$ be random variables. Then the following properties hold.
\begin{itemize}
	\item[$P1$.][Symmetry] \ind{X}{Y}{Z} $\Rightarrow$ \ind{Y}{X}{Z}.
	\item[$P2$.] \ind{X}{Y}{X}.
	\item[$P3$.][Decomposition] \ind{X}{Y}{Z} and $W \preceq Y$ $\Rightarrow$ \ind{X}{W}{Z}.
	\item[$P4$.][Weak Union] \ind{X}{Y}{Z} and $W \preceq Y$ $\Rightarrow$ \ind{X}{Y}{(W,Z)}.
	\item[$P5$.][Contraction] \ind{X}{Y}{Z} and \ind{X}{W}{(Y,Z)} $\Rightarrow$ \ind{X}{(Y,W)}{Z}.
\end{itemize}
\end{theorem}

\thmref{axioms}, stated above for stochastic variables, also holds  under conditions when some of the variables involved are non-stochastic and conditional independence is defined as in \defref{nstoc1} and \defref{nstoc2} \citep{con:14}. More precisely, suppose that we are given a collection of extended conditional independence properties. Any deduction made using the axioms of stochastic conditional independence will be valid, so long as, in both premises and conclusions, no non-stochastic variables appear in the left-most term in a conditional independence statement and one of the following conditions holds: the regime space is discrete or the stochastic variables involved are discrete or there exists a dominating regime. In our context, we only consider a finite regime space ($\cS=\{0,1,s\}$ or $\cS=\{0,1,f\}$) which validates the use of \thmref{axioms} for stochastic and non-stochastic variables together. 

\section{Identification of the Average Causal Effect at the Threshold}
\label{sec:4}
To compute the $ACE_{x_0}$, we explore conditions that allow us to identify the conditional expectations $\mE_1(Y \mid X=x_{0})$ and $\mE_0(Y \mid X=x_{0})$ from observational data. Before we examine the sharp and fuzzy RDD separately, we assume that the following two properties (\eqref{1} and \eqref{2}) hold regardless of which is the operating observational regime. 
~\\
\begin{condition}[Sufficiency]
\label{cond:suffcov}
\begin{subequations} 
\begin{align}
& \indo{(C,X)}{\Sigma} \label{eq:1} \\
& \ind{Y}{\Sigma}{(C,X,T)} \label{eq:2}
\end{align}
\end{subequations}
\end{condition}

Variables $(C,X)$ which satisfy \eqref{1} are called \textit{covariates} and variables $(C,X)$ which satisfy \eqref{1} and \eqref{2} are called {\em sufficient covariates} \citep{guodaw:10}. While the formal definitions for these properties follow from \defref{nstoc2} (for either $\cS=\{s,0,1\}$ or $\cS=\{f,0,1\}$), here we emphasize on the intuitive understanding. Property \eqref{1} requires that the distribution of $(C,X)$ is the same in all regimes. We consider this assumption appropriate as $(C,X)$ represent attributes of the individuals which are determined independently of \textit{how} the treatment is allocated. Thus the joint distribution of $(C,X)$ is considered independent of the regime $\Sigma$. Property \eqref{2} requires that the conditional distribution of $Y$ given $(C,X,T)$ also does not depend on the regime. Informally, this means that conditioning on $(C,X,T)$ we do not need further information on the regime $\Sigma$ to make probabilistic inference about $Y$. Property \eqref{2} is problem-specific as there may be several (distinct) choices for $C$, or none at all that validate this property. It lies on the researcher to identify appropriate (possibly multivariate) covariates $C$ so that \eqref{2} holds, but without it further progress seems impossible. Property \eqref{2} has also been described as `strongly ignorable treatment assignment, given U' \citep{rosrub:83}. 

Conditional independence properties \eqref{1} and \eqref{2} can be represented graphically (see Figure \ref{sup_dag1}). In general, we shall present corresponding influence diagrams (when such diagrams exist) but, for a detailed account of the semantics of influence diagrams and how they can be used to graphically derive further conditional independence properties, the reader is referred to \citet{daw:02} and \citet{cow:07}.  We note that graphical representation, when available, may be helpful as it provides a visually transparent way of deriving further conditional independence properties. However, it is never essential and all that can be achieved using the graphical approach, and more, can be achieved using an algebraic approach. 

\subsection{Sharp Design}

The property that characterises the sharp RDD and separates it from the fuzzy RDD is that the guidelines are strictly adhered to. This property can be expressed mathematically as follows.  
~\\
\begin{property}
\label{prop:s1a}
\begin{equation}
\label{eq:s1a}
\ind{T}{C}{(X, \Sigma=s)}
\end{equation}
\end{property}

\propref{s1a} states that in the observational regime $\Sigma=s$, $T$ does not depend on $C$, given $X$. That is, the only information that is used to decide on the treatment $T$, is that provided by the assignment variable $X$, and no further information that might be contained in $C$ is relevant. This property is trivially true for the interventional regimes as well. Once we condition on the value of the interventional regime, the treatment takes the same value as the value of the regime with probability one and further information on any other variable becomes irrelevant. Thus, when the observational regime represents a sharp RDD and $\cS=\{0,1,s\}$, \condref{s1} below holds.
~\\
\begin{condition}
\label{cond:s1}
\begin{equation}
\label{eq:3}
\ind{T}{C}{(X, \Sigma)}
\end{equation}
\end{condition}

We can represent \eqref{1}, \eqref{2} and \eqref{3} graphically using an influence diagram (see Figure \ref{sup_dag2}). Combining \condref{suffcov} and \condref{s1}, we can show that \ind{Y}{\Sigma}{(X,T)}.
~\\
\begin{theorem}
\label{thm:s1}
Let $(C,X)$ be sufficient covariates and assume that \condref{s1} holds. Then,
\begin{equation}
\label{eq:s1}
\ind{Y}{\Sigma}{(X,T)}.
\end{equation}
\end{theorem}

\thmref{s1} implies that the outcome of interest $Y$, no longer depends on the regime $\Sigma$, once we have information on the assignment variable $X$ and the treatment variable $T$. What is important to notice about this result, is that it allows us to transfer probabilistic information about $Y$, between the regimes, completely disregarding the extra information that is provided by $C$. This result (see \thmref{s2}), allows us to identify the $ACE_{x_0}$ from the observational sharp RDD. 

The last condition that is required, in order to justify estimation of the $ACE_{x_0}$ from observations {\emph{around} the threshold (as opposed to observations {\emph at} the threshold), is the continuity assumption, expressed in \condref{s2}. 
~\\
\begin{condition}[Continuity in the interventional regimes] 
\label{cond:s2} 
\begin{subequations}
\begin{align} 
& \mE_{0}(Y \mid X=x_{0})= \lim_{\epsilon \downarrow 0}\mE_{0}(Y \mid X=x_{0}-\epsilon) \label{eq:s2a}\\
& \mE_{1}(Y \mid X=x_{0})= \lim_{\epsilon \downarrow 0}\mE_{1}(Y \mid X=x_{0}+\epsilon) \label{eq:s2b} 
\end{align}
\end{subequations}
\end{condition}

\condref{s2} requires that the expectation of the variable of interest $Y$ is left (right) continuous at the threshold $x_{0}$, in the interventional regime $\Sigma=0$ ($\Sigma=1$). Violation of this assumption would suggest that a change in the expectation of $Y$ is due to a change in $X$ (around the threshold $x_{0}$) and not necessarily a change in $T$. Analogues of this assumption can also be found in the counterfactual framework \citep{hahtod:01, dav:08}.
~\\
\begin{theorem} 
\label{thm:s2}
Let $(C,X)$ be sufficient covariates and assume that \condref{s1} and \condref{s2} hold. Then,
\begin{equation*}
\label{eq:sid}
ACE_{x_0}= \lim_{\epsilon \downarrow 0}[\mE_{s}(Y \mid X=x_0+\epsilon, T=1)- \mE_{s}(Y \mid X=x_0-\epsilon, T=0)]. 
\end{equation*}
\end{theorem}
This theorem shall be used to prove \thmref{s3} (see supporting material).

The RDD has been linked with the instrumental variables (IV) framework as the threshold indicator $Z$ satisfies the properties of a binary IV \citep{imblem:08, didmenshe:10, gen:15}. For the sharp RDD, it appears that we do not need to use the properties of an IV, in order to prove an alternative identification formula based on the threshold indicator $Z$.
~\\
\begin{theorem}
\label{thm:s3}
Let $(C,X)$ be sufficient covariates and assume that \condref{s1} and \condref{s2} hold. Then,
\begin{equation*}
\label{eq:sidiv}
ACE_{x_0}= \lim_{\epsilon \downarrow 0}[\mE_{s}(Y \mid X=x_0+\epsilon, Z=1)-\mE_{s}(Y \mid X=x_0-\epsilon, Z=0)]. 
\end{equation*}
\end{theorem}
Hence, the $ACE_{x_0}$ is identified under a sharp RDD. We now turn our attention to the, more common, fuzzy RDD.
\subsection{Fuzzy Design}

In the fuzzy RDD, in addition to information contained in $X$, further information contained in $C$ is taken into account in order to determine the treatment $T$. Thus \nind{T}{C}{(X,\Sigma=f)} and, as a consequence, we cannot derive \eqref{s1} which allows us to identify the $ACE_{x_{0}}$ from the observational regime. In this section, we explore different conditions which allow identification of the $ACE_{x_{0}}$ from a fuzzy RDD.
~\\
\begin{condition}
\label{cond:f1}
For $t=0,1$,
\begin{equation}
\label{eq:f1}
\mP_{f}(T=t \mid C, X)>0 \quad \as{\mP_{f}} 
\end{equation}
\end{condition}

Variables $(C,X)$ which satisfy \eqref{1}, \eqref{2} and \eqref{f1} are called {\em strongly sufficient covariates} \citep{guodaw:10}. \condref{f1} states that, given any observable values of $(C,X)$, there is positive probability to observe both active and control treatment in the observational regime. For values $(C,X) = (c,x)$ around the threshold $X=x_{0}$, we expect this condition to hold as both treatments will be observed with high probability. Here we also require that both treatments are observed with probability greater than zero for values $(C,X) = (c,x)$  away from the threshold $X=x_{0}$. 
~\\
\begin{condition}[Continuity in the observational regime]
\label{cond:f2}
For $t=0,1$, the conditional expectation $\mE_{f}(Y \mid C, X , T=t)$ is continuous in $x$ at $x_{0}$.
\end{condition}

\condref{f2} requires that for given values of the treatment $T$ (active and control), the conditional expectation of $Y$, given $C$, $X$ and $T$ is continuous at the threshold $x_{0}$. This condition suggests that it is the treatment and not any of the other variables that is responsible for the discontinuity in the outcome around the threshold. Similar assumptions can be found in \citep{hahtod:01, dav:08, gen:15}.

~\\
\begin{theorem}
\label{thm:f1}
Let $(C,X)$ be strongly sufficient covariates. Then for $t = 0,1$ and any versions of the conditional expectations,
\begin{equation}
\label{eq:thm1}
\mE_{t}(Y \mid C,X) = \mE_{f}(Y \mid C, X, T = t) 
\end{equation}
almost surely in any regime. In particular, 
\begin{equation}
\label{eq:thm1b}
\mE_{t}(Y \mid C,X) = \mE_{f}(Y \mid C, X, T = t) \quad \as{\mP_{f}}.
\end{equation}
\end{theorem}

~\\
\begin{theorem}
\label{thm:f2}
Let $(C,X)$ be strongly sufficient covariates and assume that \condref{f2} holds. Then,
\begin{equation*}
\label{eq:fid1a}
ACE_{x_0}= \mE_{f}[\lim_{\epsilon \downarrow 0} h(C) \mid X=x_{0}]
\end{equation*}
where 
\begin{equation*}
\label{eq:fid1b}
h(C)= \mE_{f}(Y \mid C, X=x_0+\epsilon, T=1)- \mE_{f}(Y \mid C, X=x_0-\epsilon, T=0). 
\end{equation*}

\end{theorem}

\thmref{f2} provides a formula for identifying the $ACE_{x_0}$ purely from observational data as we have shown that the $ACE_{x_0}$ is expressible purely in terms of properties of the observational joint distribution of $(Y, C, X, T)$, where $(C, X)$ are strongly sufficient covariates.

To identify the $ACE_{x_{0}}$ in terms of the threshold indicator $Z$ which is considered a special case of a binary IV \citep{imblem:08, didmenshe:10, gen:15}, we invoke the following IV property for $\cS=\{0, 1, f\}$.
~\\
\begin{property}[IV property]
\label{prop:f3}
\begin{equation}
\label{eq:f3}
\ind{Y}{Z}{(C,X,T,\Sigma)}
\end{equation}
\end{property}
\propref{f3} requires that the conditional distribution of $Y$ given $(C, X, T)$ in any regime $\sigma \in \cS=\{0, 1, f\}$ does not depend on $Z$. This property readily follows as $Z$ is a function of $X$ and, once $X$ is given, further information about $Z$ becomes redundant for $Y$.  

\begin{lemma}
\label{lem:f1}
Assume that \condref{f2} and \propref{f3} hold. Then  \as{\mP_{f}},
\begin{align*}
 & \lim_{\epsilon \downarrow 0}[\mE_{f}(Y \mid C, X=x_{0}+\epsilon, T=1) -\mE_{f}(Y \mid C, X=x_{0}-\epsilon, T=0)] \\
 & =\lim_{\epsilon \downarrow 0} \frac{\mE_{f}(Y \mid C, X=x_{0}+\epsilon, Z=1) \mE_{f}(Y \mid C, X=x_{0}-\epsilon, Z=0)}{\mP_{f}(T=1 \mid C, X=x_{0}+\epsilon, Z=1) - \mP_{f}(T=1 \mid C, X=x_{0}-\epsilon, Z=0)}.
\end{align*}
\end{lemma}
Lemma \ref{lem:f1} leads to the following Theorem, in  which the average causal effect is identified.
~\\
\begin{theorem}
\label{thm:f4}
Let $(C,X)$ be strongly sufficient covariates and assume that \condref{f2} and \propref{f3} hold. Then,
\begin{equation*}
\label{eq:fid2a}
ACE_{x_0}= \mE_{f}[\lim_{\epsilon \downarrow 0} h(C) \mid X=x_{0}]
\end{equation*}
where 
\begin{equation*}
\label{eq:fid2b}
h(C)= \frac{\mE_{f}(Y \mid C, X=x_0+\epsilon, Z=1)- \mE_{f}(Y \mid C, X=x_0-\epsilon, Z=0)}{\mP_{f}(T=1 \mid C, X=x_{0}+\epsilon, Z=1)- \mP_{f}(T=1 \mid C, X=x_{0}-\epsilon, Z=0)}. 
\end{equation*}
\end{theorem}

In the next section, we present a worked example in which the methods described are applied to real example concerning the prescription of statins in UK primary care.

\section{Example: Prescription of Statins in UK Primary Care}
\label{sec:5}
We describe an example in which the methods outlined in Sections 2--4 are applied to a set of real data. We use The Health Improvement Network (THIN) database, a large source of UK primary care data, consisting of routine, anonymised, patient records collected during patient consultations at over 500 general practices (GPs)  (www.epic-uk.org). We wish to  examine the causal relationship between the initiation of statin therapy and the low density lipoprotein (LDL) cholesterol level. Many randomised trials and other studies have established the effect of statins on LDL cholesterol level (\cite{sca:94}, \cite{hps:02}, \cite{bai:05}, \cite{bru:09}) which makes this choice of example useful for demonstrative purposes.  

In January 2006, the UK National Institute of Health and Care Excellence (NICE) issued guidelines stating that individuals whose 10-year risk of cardiovascular disease (CVD) development was greater than 20\% should be prescribed statins, a class of cholesterol-lowering drugs (\cite{NICE}). Here, 10-year CVD risk is measured using an appropriate risk predictor, such as the 10-year Framingham risk score (\cite{wil:98}) or Q-RISK score (\cite{hip:07}). We consider an individual's 10-year CVD risk score as the `assignment variable' for statin therapy. 

We chose to use data from 1386 male patients in THIN, for whom a 10-year CVD risk was calculated between 01 January 2007 and 31 December 2008, aged 50--70 who were non-diabetic, non-smokers, had not previously experienced a cardiovascular event (e.g.~stroke, myocardial infarction) and who had not previously been prescribed statins. We define three key variables of interest in this example, where $i \in\{1,\ldots,1386\}$ denotes the patient index:
\begin{itemize}
\item{$X_{i} \in[0,1]$: Risk score of the $i^{\text{th}}$ patient (the \textbf{assignment variable}).}
\item{$Y_{i}\in[0,\infty)$: LDL cholestrol level of the $i^{\text{th}}$ patient (the \textbf{continuous outcome variable}).}
\item{$T_{i}\in\{0, 1\}$: Treatment indicator for the $i^{\text{th}}$ patient (= 0 if patient is not prescribed statins, = 1 if patient is prescribed statins).}\end{itemize}
We note that $Y_{i}$ is measured between one and six months after the risk score calculation, so that the effect of statins on LDL cholesterol level can be observed. Furthermore, we define the threshold attainment indicator $Z_{i}$ ($Z_{i} = 1$ if risk score is at or above the threshold and $Z_{i} = 0$ otherwise).

Figure \ref{sca_plot} shows a simple scatter plot of the assignment variable (10-year CVD risk score) against the outcome variable (LDL cholesterol level), with different icons to differentiate between patients who did and did not receive statins. We see that, as would be expected, the RDD is fuzzy in this example. Nonetheless, there  appears to be an obvious separation in the  LDL cholesterol levels of the treated and the untreated at the treatment threshold suggesting, at first sight, that the use of a fuzzy RDD may be appropriate for these data.  

We consider the possible confounders. Since 10-year risk score is calculated based on: age, gender, smoking status, diabetic status, systolic and diastolic blood pressures and total cholesterol level, it is perhaps reasonable to consider these variables as the main confounders in this example.  The data have been stratified by choosing only males aged 50--70 who are non-diabetic and non-smokers. We assume that other confounders are similarly distributed for patients whose risk scores lie close enough to the treatment threshold. The notion of patients `lying close to the treatment threshold' is reflected in the choice of RDD bandwidth, which we denote  $\zeta$ (with $\zeta>0$). Patients' data are considered in the RDD if their assignment variable value lies within the interval $(x_{0}-\zeta, x_{0}+\zeta)$, representing the range of assignment variable values for which patients are considered to be exchangeable in the RDD. 

For most patients, we would expect this assumption of confounders being similarly distributed above and below the threshold to be reasonable, especially after stratification and where the chosen bandwidth is reasonably small, although the assumption is untestable. However, as is often seen in randomised trials, it can be useful to compare basic summary statistics of possible confounders between treatment groups.  Summary statistics for confounders compared between groups above and below the threshold are shown in Table \ref{sum_stats}, for $\zeta\in\{0.05, 0.10, 0.15\}$. These show that, in general, confounders appear to be similarly distributed in groups above and below the threshold.

\begin{table}[ht]
\begin{center}
\caption{\label{sum_stats}Table showing summary statistics for a variety of possible confounding variables for patients above and below the threshold, using a variety of RDD bandwidths ($\zeta = 0.05, 0.1$ or $0.15$), for the THIN data subset. }
\fbox{ 
 \begin{tabular}{ccccccc}
  \multicolumn{7}{l}{\emph{Bandwidth = 0.05}}\\
  Variable & Group & Mean & Median & Std. Dev. & Minimum & Maximum\\
\hline
 \multirow{2}{*}{Systolic BP (mmHg)} & $Z_{i}=0$ & 133.7 & 134.0 & 9.9 & 106.0 & 162.0\\
                                             ~                          & $Z_{i}=1$ & 139.5 & 139.0 & 11.7 & 110.0 & 184.0\\
\hline
 \multirow{2}{*}{Diastolic BP (mmHg)} & $Z_{i}=0$ & 81.6 & 80.0 & 8.3 & 59.0 & 110.0\\
                                              ~                        & $Z_{i}=1$ & 82.0 & 80.0 & 8.8 & 58.0 & 110.0\\
\hline
\multirow{2}{*}{LDL Cholesterol (mmol/L)} & $Z_{i}=0$ & 3.96 & 3.93 & 0.82 & 1.50 & 6.20\\
                                               ~                                     & $Z_{i}=1$ & 3.93 & 3.90 & 0.77 & 1.24 & 6.20\\
\hline
\multirow{2}{*}{HDL Cholesterol (mmol/L)} & $Z_{i}=0$ & 1.34 & 1.30 & 0.31 & 0.73 & 2.85\\
                                                ~                                     & $Z_{i}=1$ & 1.25 & 1.20 & 0.27 & 0.76 & 2.28\\
\hline
\multirow{2}{*}{Triglycerides (mmol/L)} & $Z_{i}=0$ & 1.68 & 1.60 & 0.78 & 0.38 & 5.10\\
                                                 ~                                    & $Z_{i}=1$ & 1.78 & 1.60 & 0.91 & 0.55 & 6.78\\
\hline
 \multicolumn{7}{l}{\emph{Bandwidth = 0.10}}\\
 Variable & Group & Mean & Median & Std. Dev. & Minimum & Maximum\\
\hline
\multirow{2}{*}{Systolic BP (mmHg)} & $Z_{i}=0$ & 131.7 & 132.0 & 10.6 & 102.0 & 176.0\\
                                     ~                                  & $Z_{i}=1$ & 141.2 & 140.0 & 12.4 & 110.0 & 199.0\\
\hline
 \multirow{2}{*}{Diastolic BP (mmHg)} & $Z_{i}=0$ & 81.1 & 80.0 & 8.1 & 59.0 & 110.0\\
                                      ~                                 & $Z_{i}=1$ & 82.5 & 81.0 & 9.1 & 58.0 & 110.0\\
\hline
\multirow{2}{*}{LDL Cholesterol (mmol/L)} & $Z_{i}=0$ & 3.82 & 3.76 & 0.82 & 0.90 & 6.20\\
                                       ~                                             & $Z_{i}=1$ & 3.99 & 3.90 & 0.80 & 1.24 & 6.90\\
\hline
\multirow{2}{*}{HDL Cholesterol (mmol/L)} & $Z_{i}=0$ & 1.37 & 1.30 & 0.32 & 0.73 & 2.85\\
                                        ~                                             & $Z_{i}=1$ & 1.24 & 1.20 & 0.26 & 0.60 & 2.28\\
\hline
\multirow{2}{*}{Triglycerides (mmol/L)} & $Z_{i}=0$ & 1.57 & 1.40 & 0.74 & 0.38 & 5.10\\
                                         ~                                            & $Z_{i}=1$ & 1.80 & 1.60 & 0.91 & 0.55 & 9.00\\
\hline
 \multicolumn{7}{l}{\emph{Bandwidth = 0.15}}\\
 Variable & Group & Mean & Median & Std. Dev. & Minimum & Maximum\\
\hline
\multirow{2}{*}{Systolic BP (mmHg)} & $Z_{i}=0$ & 130.3 & 130.0 & 11.2 & 97.0 & 176.0\\
                                        ~                               & $Z_{i}=1$ & 142.5 & 140.0 & 12.9 & 110.0 & 202.0\\
\hline
 \multirow{2}{*}{Diastolic BP (mmHg)} & $Z_{i}=0$ & 80.5 & 80.0 & 8.2 & 59.0 & 110.0\\
                                         ~                              & $Z_{i}=1$ & 83.0 & 82.0 & 9.2 & 58.0 & 119.0\\
\hline
\multirow{2}{*}{LDL Cholesterol (mmol/L)} & $Z_{i}=0$ & 3.75 & 3.70 & 0.83 & 0.90 & 6.20\\
                                         ~                                           & $Z_{i}=1$ & 4.00 & 3.90 & 0.78 & 1.24 & 6.90\\
\hline
\multirow{2}{*}{HDL Cholesterol (mmol/L)} & $Z_{i}=0$ & 1.39 & 1.33 & 0.33 & 0.73 & 2.85\\
                                         ~                                            & $Z_{i}=1$ & 1.21 & 1.20 & 0.26 & 0.60 & 2.28\\
\hline
\multirow{2}{*}{Triglycerides (mmol/L)} & $Z_{i}=0$ & 1.52 & 1.36 & 0.73 & 0.38 & 5.10\\
                                         ~                                            & $Z_{i}=1$ & 1.85 & 1.62 & 0.92 & 0.55 & 9.00\\
\end{tabular}}
\end{center}
\end{table}

The THIN data have been generated under an observational regime. In accordance with the ideas presented in Section 4, we must examine whether or not we can identify the average causal effect of statins on the LDL cholesterol level using these data. Condition (4.1a) implies that the distributions of the 10-year risk score and the confounding variables should be independent of the regime under which the data were obtained. Whilst this condition is untestable, pragmatically, it appears reasonable that the way in which treatment arises is unlikely to have affected the distributions of confounders and the 10-year CVD risk score, especially in patients whose 10-year CVD risk scores lie close to the threshold.  For Condition (4.1b) to be satisfied, we consider the further assumption that the LDL cholesterol level is independent of the regime, conditional on the confounders, 10-year CVD risk score and treatment allocation. Again, although this is untestable, it seems reasonable to assume there may be correlation between the confounders, the risk score and LDL cholesterol level and, if statins affect LDL cholesterol level, correlation between LDL cholesterol level and treatment allocation via some biological mechanism through which statins work on human subjects. Once these relationships have been accounted for, we assume that the mechanism through which the data concerning treatment (or non-treatment) arose (i.e.~interventional or observational) is of limited importance. Consequently, we argue that Condition (4.1b) is likely to be satisfied in this example.

Condition 4.8 states that the both the probability of treatment and the probability of non-treatment should be non-zero, almost surely, under the fuzzy RDD, conditional on the assignment variable and the set of confounders.  In this example, Figure \ref{sca_plot} and the results in Table \ref{sum_stats} show that the probability of treatment is non-zero under many different levels of the possible confounding variables and the 10-year risk score. Hence, we argue that the confounders and the 10-year risk score are \textit{strongly sufficient} covariates. 

Condition 4.9 states that the expected value of the outcome variable should be continuous in the assignment variable, $x$, at the threshold conditional on the confounders, 10-year risk score and where treatment does not change. In this example, it is unlikely that the expected LDL cholesterol value is likely to `jump' where treatment assignment does not change. 

Finally, we need to argue that Property 4.12 holds with these data. In essence, Property 4.12 implies that the outcome variable value is independent of the threshold variable conditional on the assignment variable, treatment, regime indicator and confounders. In this example, the  treatment rule (prescription of statins if the 10-year CVD risk score is 20\% or greater) is determined by the UK government and is not patient-specific. Furthermore, since $Z$ is determined solely by the 10-year CVD risk score, conditioning on this variable implies directly that further knowledge of $Z$ is unnecessary to make inference regarding the LDL cholesterol level. Hence, we apply the result of Theorem 4.14 to estimate the causal effect of statin therapy on LDL cholesterol level, using a fuzzy RDD and the THIN subset of data. 

\subsection{Models and Estimation}
 For a chosen bandwidth $\zeta$ ($\zeta\in(0,1)$), we define $\mathcal{A}_{\zeta}$ and $\mathcal{B}_{\zeta}$ to be the sets of patients whose 10-year CVD risk scores lie above and below the threshold, respectively. Setting the treatment threshold to be $x_{0} = 0.2$ and defining $X^{C}_{i} = X_{i}-0.2$ to be the centred 10-year CVD risk score for the $i^{\text{th}}$ patient, we consider the following linear models:
\begin{align*}
Y_{i} &= \beta_{0a}+\beta_{1a}X^{C}_{i}+\omega_{i} \text{~~(for $i\in\mathcal{A}_{\zeta}$);}\\
Y_{i} &= \beta_{0b}+\beta_{1b}X^{C}_{i}+\omega_{i} \text{~~(for $i\in\mathcal{B}_{\zeta}$).}
\end{align*}
with $\omega_{i}\sim\mathcal{N}(0,\sigma^{2})$ for $i \in\mathcal{A}_{\zeta}\cup\mathcal{B}_{\zeta}$. The function that we wish to evaluate is
\begin{equation}\label{late}
h(C)= \frac{\mE_{f}(Y \mid C, X=x_0+\epsilon, Z=1)- \mE_{f}(Y \mid C, X=x_0-\epsilon, Z=0)}{\mP_{f}(T=1 \mid C, X=x_{0}+\epsilon, Z=1)- \mP_{f}(T=1 \mid C, X=x_{0}-\epsilon, Z=0)}.
\end{equation}
The numerator of (\ref{late}) is estimated by $\hat{\beta}_{0a}-\hat{\beta}_{0b}$. To estimate the denominator of (\ref{late}) we consider the following estimates:
\begin{align*}
\mP_{f}(T=1 \mid C, X=x_{0}+\epsilon, Z=1) &\approx \hat{\pi}_{a} =  \frac{1}{|\mathcal{A}_{\zeta}|}\sum_{i\in\mathcal{A}_{\zeta}}T_{i}\\
\mP_{f}(T=1 \mid C, X=x_{0}-\epsilon, Z=0) &\approx  \hat{\pi}_{b} = \frac{1}{|\mathcal{B}_{\zeta}|}\sum_{i\in\mathcal{B}_{\zeta}}T_{i}
\end{align*}
Hence, the causal effect estimate for the effect of statins on LDL cholesterol level is given by
\[
\hat{\eta} = \frac{\hat{\beta}_{0a}-\hat{\beta}_{0b}}{\hat{\pi}_{a}-\hat{\pi}_{b}}
\]
where $(\hat{\beta}_{0a}, \hat{\beta}_{0b})$ denote the maximum likelihood estimates of $(\beta_{0a}, \beta_{0b})$. We consider the calculation of $\eta$ using the THIN subset data and a choice of three RDD bandwidths ($\zeta = 0.05, 0.1$ and $0.15$). Estimated values, $\hat{\eta}$, together with associated 95\% confidence intervals, are shown in Table \ref{LATE_res}. 

Examining Table \ref{LATE_res}, we see that for each chosen RDD bandwidth, the estimated causal effect of statins on LDL cholesterol level appears to differ from zero, suggesting that statins appear to reduce the LDL cholesterol level in general. This would be expected, especially in light of the many large-scale randomised trials that have provided substantial evidence in favour of the beneficial effect of statins with respect to LDL cholesterol level. We have used the arguments and results presented in earlier sections of the paper to justify and construct a regression discontinuity design on real observational data, through a decision theoretic approach. We note that the use of a restricted sample of patients for this demonstrative example implies that the results reported are not necessarily applicable to the general population, or of strict clinical/epidemiological significance.

\begin{table}
\begin{center}
 \caption{\label{LATE_res} Table showing the estimated causal effect of statin therapy on LDL cholesterol level, together with associated standard errors and 95\% confidence intervals using the THIN data subset for chosen bandwidths: 0.05, 0.10 and 0.15.}
   \fbox{%
   \begin{tabular}{ccc}
    Bandwidth ($\zeta$) & Estimate $\hat{\eta}$ (Standard Error) & 95\% Confidence Interval\\
  \hline
  0.05 & -0.766 (0.373) & (-1.497, -0.035)\\
  0.10 & -0.889 (0.209) & (-1.299, -0.479)\\
  0.15 & -0.934 (0.165) & (-1.258, -0.610)\\ 
\end{tabular}}
 \end{center}
 \end{table}

\section{Discussion}
\label{sec:6}
We have developed a formal, decision theoretic approach to the regression discontinuity design using the language of conditional independence. Conditions under which the causal effect of a treatment on a continuous outcome of interest can be identified were defined and the corresponding causal effect estimator derived, for both sharp and fuzzy RDDs. Full theoretical justification for causal effect identification was made. 

In line with other methods for causal inference in observational studies, assumptions involving confounders are necessary, though we argue that, where such assumptions are reasonable, carefully-argued and example-specific, estimation of a causal effect can be made in either a sharp or fuzzy RDD. Specification of a full and appropriate set of confounders remains important, as with other causal inference methods. With an RDD, the comparison of groups above and below the treatment threshold, but whose assignment variable values lie close to the treatment threshold, aims to ensure that the distributions of confounders, both observed and unobserved, are balanced between comparative groups. This assumption highlights the link between the RDD in an observational setting and the randomised controlled trial in an experimental setting. Although this assumption is untestable, it may seem reasonable in many scenarios and simple methods, such as basic plots and summary statistics, can be helpful in highlighting fundamental distributional discrepancies in confounders between groups above and below the treatment threshold. 

We applied the methodology to an example based on real data concerning the prescription of statins to patients at moderate risk of cardiovascular disease development, according to a 10-year CVD risk score. In this example, the main observed confounders were specified and values compared between groups above and below the treatment threshold. A basic scatter plot of the assignment variable against the outcome of interest (Figure \ref{sca_plot}) suggested that the use of a fuzzy RDD was likely to be appropriate for these data. It is important to note the value of plots such as Figure \ref{sca_plot}, which should always be produced whenever the use of an RDD is considered for any observational dataset. Justification of each assumption required in our decision theoretic approach was made and the causal effect of statins on LDL cholesterol level was calculated using a selection of pre-specified bandwidths. This example showcases the use of a rigorous approach to RDD analyses in medical studies and epidemiology. 

Further methodological extensions to our work include the development of RDD methods for non-continuous outcomes and dynamic processes. We encourage the wider use of the RDD as a valid method for treatment effect estimation in biostatistics and epidemiology. We hope that the use of RDD methods will become more widespread in this field, particularly with the increasing availability of large-scale electronic healthcare databases. 

\section*{Acknowledgements}
We thank Vanessa Didelez for helpful and insightful discussions concerning this work. Panayiota Constantinou received funding from the UK Engineering and Physical Sciences Research Council (EPSRC grant: D063485/1). Aidan O'Keeffe received funding from the UK Medical Research Council (MRC grant: MR/K014838/1). Approval for the use of the THIN data was granted by a Scientific Review Committee in August 2014 (ref.~14-021). 

\bibliography{rdd}
\bibliographystyle{apalike}

\newpage

\begin{figure}
 \begin{center}
  \includegraphics[scale=0.6]{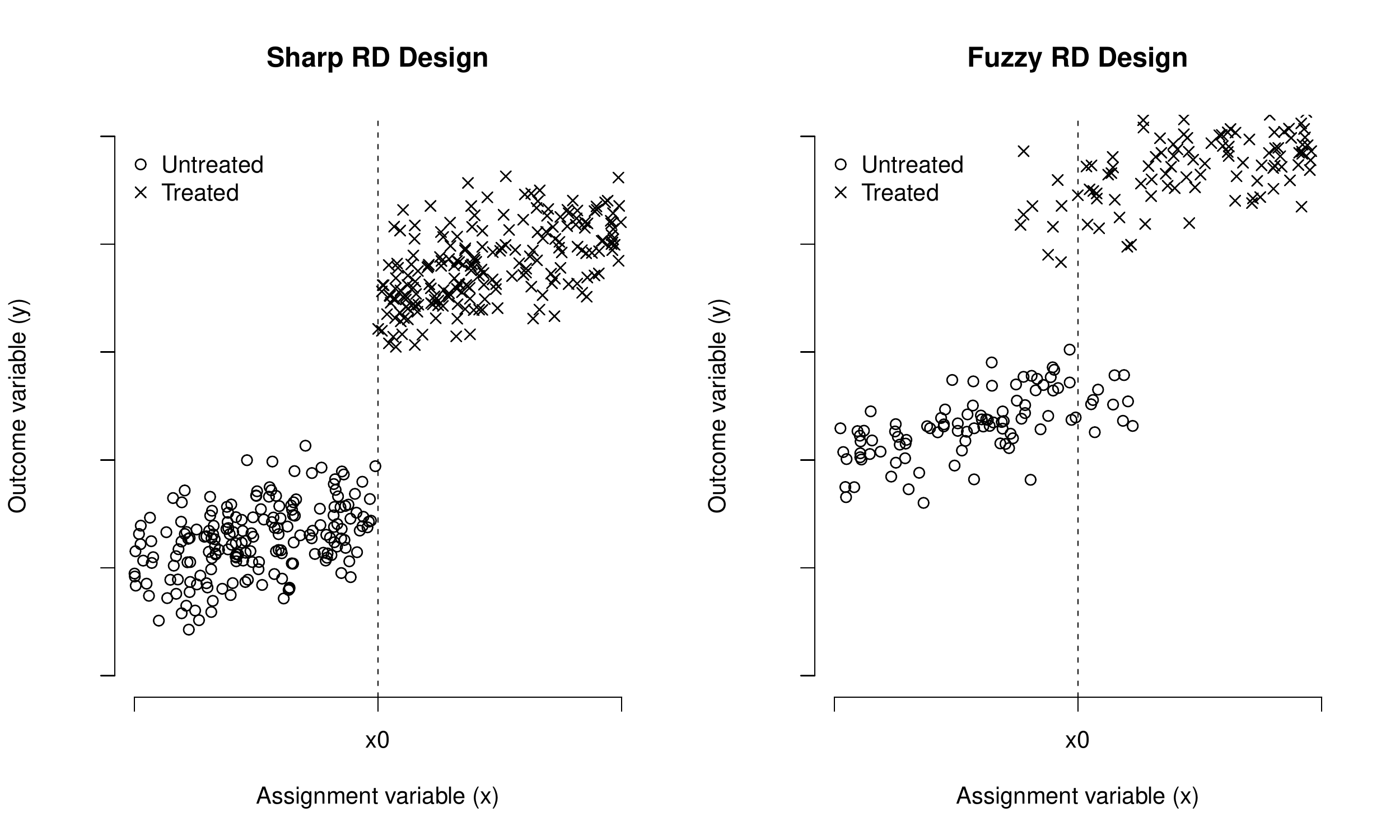}
	\caption{\label{plot1}Graphical representations of example sharp (left-hand plot) and fuzzy (right-hand plot) regression discontinuity designs. `Untreated' icons denote subjects who do not receive the treatment whereas `Treated' icons denote subjects who received the treatment. The treatment threshold is depicted by a dashed line at $x = x_{0}$.}
 \end{center}
\end{figure}

\begin{center}
\begin{figure}[ht]
\centering
\rotatebox{0}{
\scalebox{0.25}{
\includegraphics{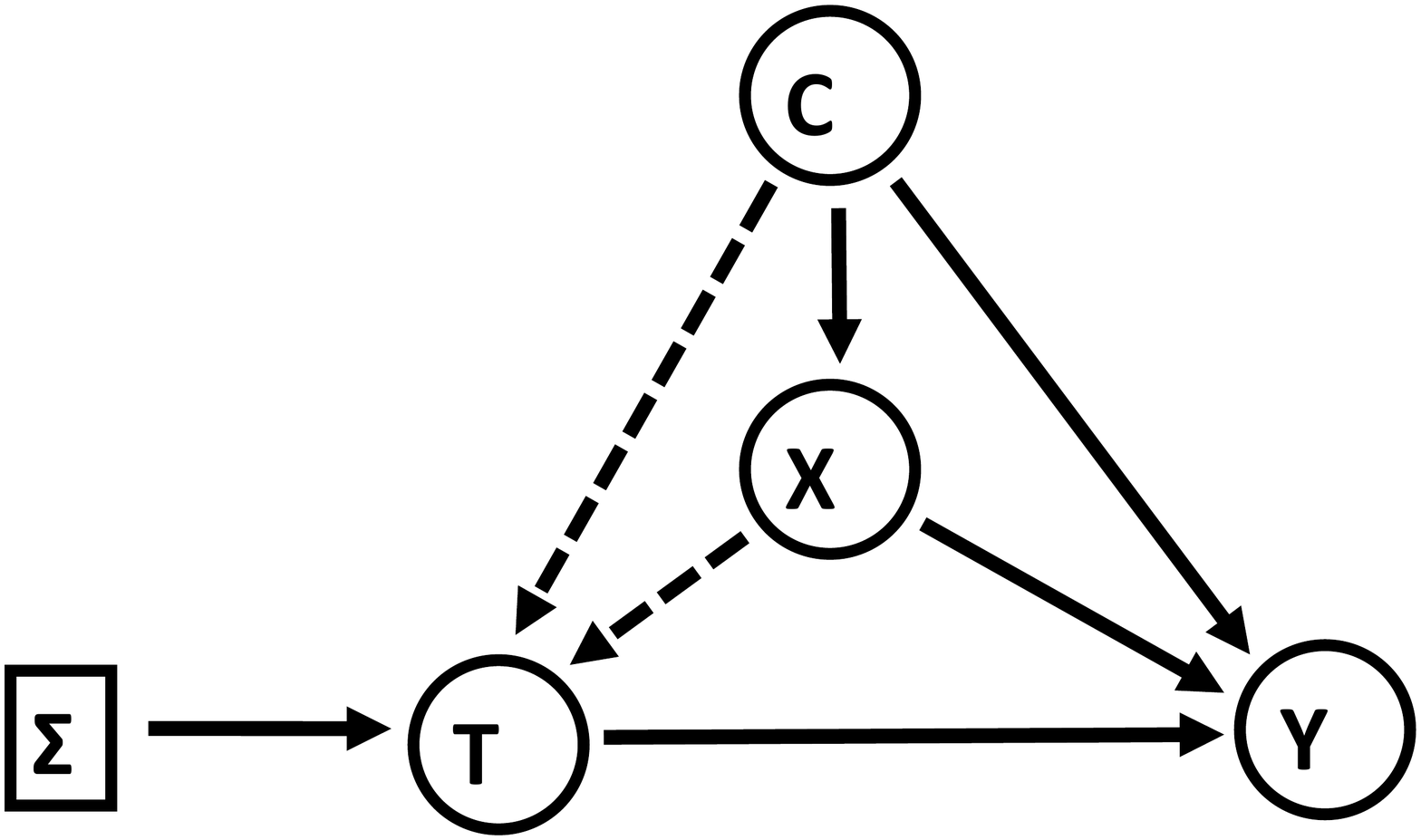}}}
\caption{\label{sup_dag1}Influence Diagram representing (4a) and (4b). Property (4a) is represented by the absence of arrows from $\Sigma$ to $(C,X)$ and property (4b) is represented by the absence of a direct arrow from $\Sigma$ to $Y$. The dotted arrows from $X$ to $T$ and from $C$ to $T$ indicate a link that disappears under an interventional regime. }
\label{fig:suffcov} 
\end{figure}
\end{center}

\newpage

\begin{center}
\begin{figure}[ht]
\centering
\rotatebox{0}{
\scalebox{0.25}{
\includegraphics{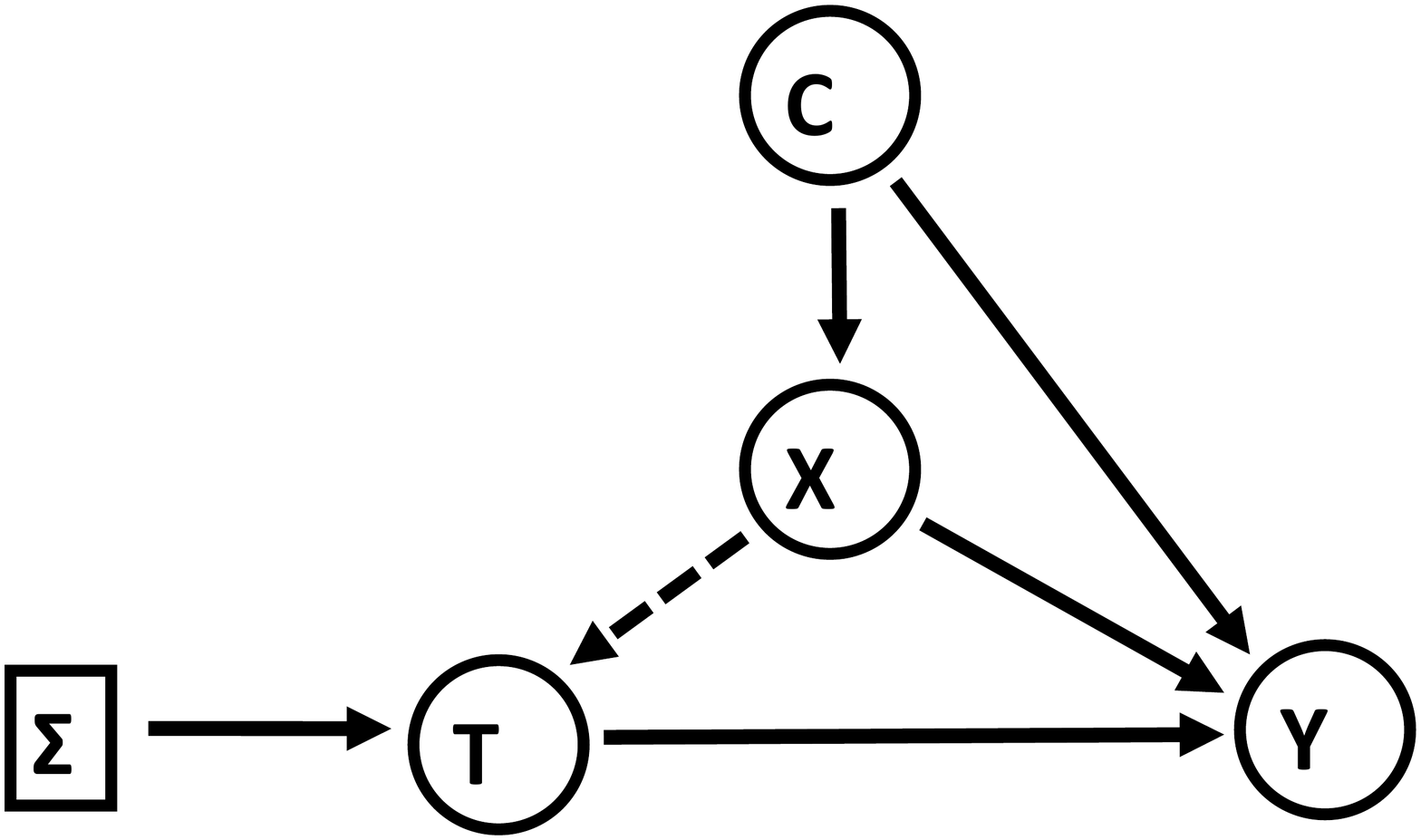}}}
\caption{\label{sup_dag2}Influence Diagram representing (4a), (4b) and (6). Property (6) is represented by the absence of an arrow  from $C$ to $T$.}
\label{fig:suffcovsharp} 
\end{figure}
\end{center}

\begin{figure}[ht]
\begin{center}
  \includegraphics[scale=0.7]{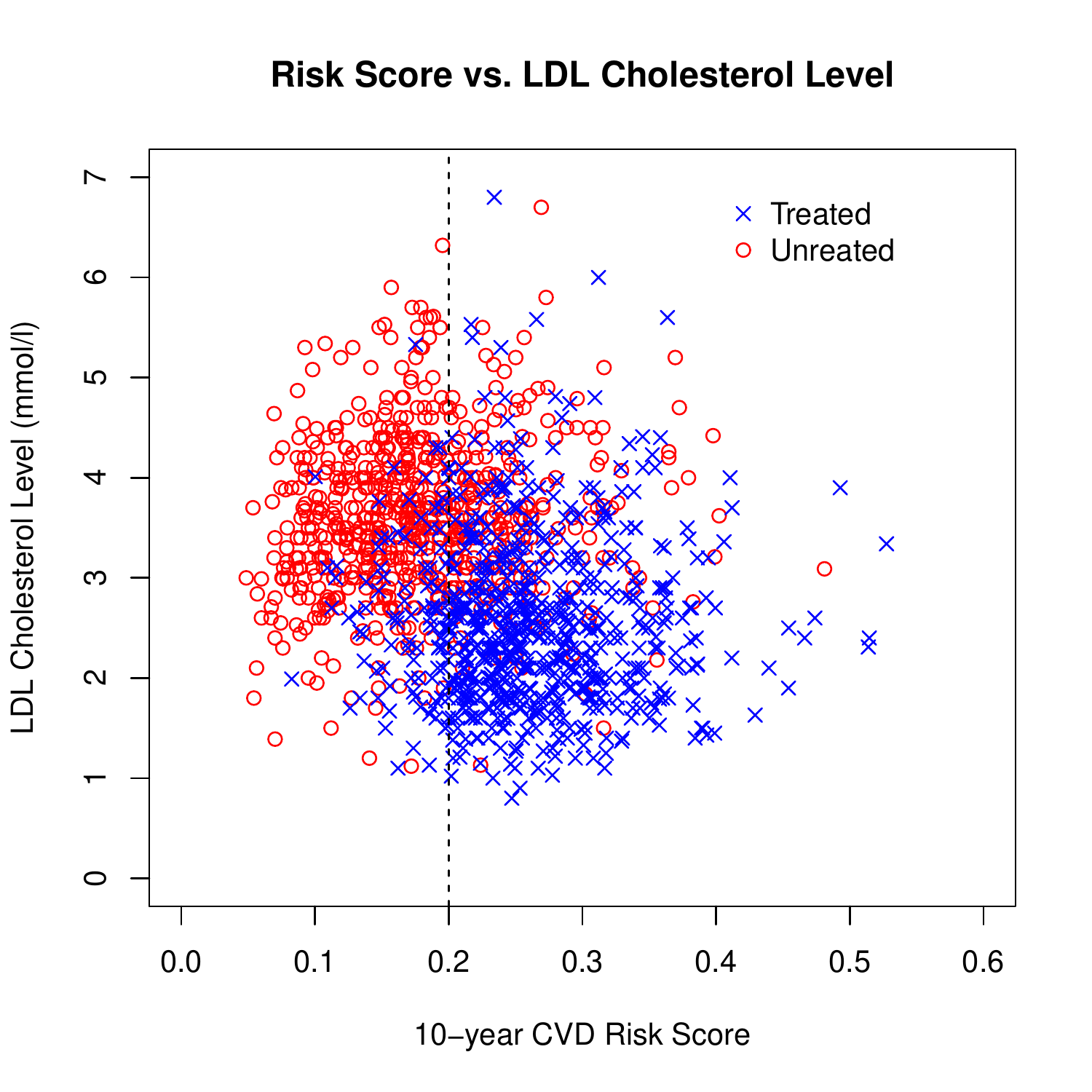}
\caption{\label{sca_plot}  Scatter plot showing 10-year CVD risk score against LDL cholesterol level for the THIN subset data. Patients who receive statins are indicated by blue icons and patients who do not receive statins are indicated by red icons. The dashed vertical line denotes the treatment threshold at the 20\% risk score.}
\end{center}
\end{figure}

\clearpage

\begin{appendix}
\section{Proofs of Theorems in Section 4}

\subsection{Proof of Theorem 4.4}

\textbf{Theorem 4.4}. 
\textit{~Let $(C,X)$ be sufficient covariates and assume that Condition 4.3 holds. Then,}
\begin{equation}
\ind{Y}{\Sigma}{(X,T)}.
\end{equation}

\begin{proof}
Applying the axioms of Theorem 3.3, we obtain:
\begin{align}
(4a) & \xRightarrow{P1} \indo{\Sigma}{(C,X)} \nonumber \\
         & \xRightarrow{P4} \ind{\Sigma}{(C,X)}{X} \nonumber \\
         & \xRightarrow{P3} \ind{\Sigma}{C}{X} \nonumber \\
         & \xRightarrow{P1} \ind{C}{\Sigma}{X}. \label{eq:sp1}
\end{align}
Also, 
\begin{equation} 
(6) \xRightarrow{P1} \ind{C}{T}{(X,\Sigma)} \label{eq:sp2},
\end{equation}
and applying $P5$ to \eqref{sp1} and \eqref{sp2}, we see that $\ind{C}{(T, \Sigma)}{X}$. Then applying in turn $P4$, $P3$ and $P1$ we obtain 
\begin{equation}
\ind{\Sigma}{C}{(X,T)} \label{eq:sp3}.
\end{equation}
Lastly,
\begin{equation} 
(4b) \xRightarrow{P1} \ind{\Sigma}{Y}{(C,X,T)} \label{eq:sp4}
\end{equation}
and applying $P5$ to \eqref{sp3} and \eqref{sp4} we see that $\ind{\Sigma}{(C,Y)}{(X,T)}$. Applying in turn $P3$ and $P1$ concludes the proof. 
\end{proof}

We note that the result of Theorem 4.4, can also be verified using graphical techniques (in particular, the d-separation \citep{pea:86,ver:90}, or the equivalent moralization criterion \citep{lau:90}).

\subsection{Proof of Theorem 4.6}
\textbf{Theorem 4.6}
\textit{Let $(C,X)$ be sufficient covariates and assume that Condition 4.3 and Condition 4.5 hold. Then,}
\begin{equation*}
ACE_{x_0}= \lim_{\epsilon \downarrow 0}[\mE_{s}(Y \mid X=x_0+\epsilon, T=1)- \mE_{s}(Y \mid X=x_0-\epsilon, T=0)]. 
\end{equation*}

\begin{proof}
\begin{align*}
ACE_{x_{0}} & := \mE_{1}(Y \mid X=x_{0})- \mE_{0}(Y \mid X=x_{0}) \\
  				  & = \lim_{\epsilon \downarrow 0} [\mE_{1}(Y \mid X=x_{0} + \epsilon)- \mE_{0}(Y \mid X=x_{0}-\epsilon)] \quad \text{(Condition 4.5)} \\
     				& = \lim_{\epsilon \downarrow 0} [\mE_{1}(Y \mid X=x_{0} + \epsilon, T=1)- \mE_{0}(Y \mid X=x_{0}-\epsilon, T=0)] \\
     				& = \lim_{\epsilon \downarrow 0} [\mE_{s}(Y \mid X=x_{0} + \epsilon, T=1)- \mE_{s}(Y \mid X=x_{0}-\epsilon,T=0)] \quad \text{(Theorem 4.4)} 
\end{align*}
\end{proof}

\subsection{Proof of Theorem 4.7}
\textbf{Theorem 4.7}
\textit{Let $(C,X)$ be sufficient covariates and assume that Condition 4.3 and Condition 4.5 hold. Then,}
\begin{equation*}
ACE_{x_0}= \lim_{\epsilon \downarrow 0}[\mE_{s}(Y \mid X=x_0+\epsilon, Z=1)-\mE_{s}(Y \mid X=x_0-\epsilon, Z=0)]. 
\end{equation*}

\begin{proof}
Noting that for the sharp observational regime, $\mP_{s}(T=Z)=1$, the proof follows directly from Theorem 4.6. 
\end{proof}

\subsection{Proof of Theorem 4.10}
\textbf{Theorem 4.10}
\textit{Let $(C,X)$ be strongly sufficient covariates. Then for $t = 0,1$ and any versions of the conditional expectations,}
\begin{equation}
\mE_{t}(Y \mid C,X) = \mE_{f}(Y \mid C, X, T = t) 
\end{equation}
almost surely in any regime. In particular, 
\begin{equation}
\mE_{t}(Y \mid C,X) = \mE_{f}(Y \mid C, X, T = t) \quad \as{\mP_{f}}.
\end{equation}

\begin{proof}
See Theorem 1 in \citet{guodaw:10}.
\end{proof}

\subsection{Proof of Theorem 4.11}
\textbf{Theorem 4.11}
\textit{Let $(C,X)$ be strongly sufficient covariates and assume that Condition 4.9 holds. Then,}
\begin{equation*}
ACE_{x_0}= \mE_{f}[\lim_{\epsilon \downarrow 0} h(C) \mid X=x_{0}]
\end{equation*}
\textit{where}
\begin{equation*}
h(C)= \mE_{f}(Y \mid C, X=x_0+\epsilon, T=1)- \mE_{f}(Y \mid C, X=x_0-\epsilon, T=0). 
\end{equation*}

\begin{proof}
\begin{align*}
ACE_{x_{0}} & := \mE_{1}(Y \mid X=x_{0})- \mE_{0}(Y \mid X=x_{0}) \\
            & = \mE_{1}[\mE_{1}(Y \mid C, X=x_{0}) \mid X=x_{0}]- \mE_{0}[\mE_{0}(Y \mid C, X=x_{0}) \mid X=x_{0}]\\
  				  & = \mE_{f}[\mE_{1}(Y \mid C, X=x_{0}) \mid X=x_{0}]- \mE_{f}[\mE_{0}(Y \mid C, X=x_{0}) \mid X=x_{0}]\text{~~(4a)} \\
     			  & = \mE_{f}[\mE_{f}(Y \mid C, X=x_{0}, T=1) -\mE_{f}(Y \mid C, X=x_{0}, T=0) \mid X=x_{0}]\\ 
						&~~\quad \text{(Theorem 4.10)} \\
     		  	& = \mE_{f}[\lim_{\epsilon \downarrow 0} h(C)  \mid X=x_{0}] \quad \text{(Condition 4.9)}
\end{align*}
\end{proof}

\subsection{Proof of Lemma 4.13}
\textbf{Lemma 4.13}
\textit{Assume that Condition 4.9 and Property 4.12 hold. Then  \as{\mP_{f}},}
\begin{align*}
 & \lim_{\epsilon \downarrow 0}[\mE_{f}(Y \mid C, X=x_{0}+\epsilon, T=1) -\mE_{f}(Y \mid C, X=x_{0}-\epsilon, T=0)] \\
 & =\lim_{\epsilon \downarrow 0} \frac{\mE_{f}(Y \mid C, X=x_{0}+\epsilon, Z=1) \mE_{f}(Y \mid C, X=x_{0}-\epsilon, Z=0)}{\mP_{f}(T=1 \mid C, X=x_{0}+\epsilon, Z=1) - \mP_{f}(T=1 \mid C, X=x_{0}-\epsilon, Z=0)}.
\end{align*}

\begin{proof}
\begin{align*}
 & \lim_{\epsilon \downarrow 0}[\mE_{f}(Y \mid C, X=x_{0}+\epsilon, Z=1) -\mE_{f}(Y \mid C, X=x_{0}-\epsilon, Z=0)] \\
 & = \lim_{\epsilon \downarrow 0} [\mE_{f}(Y \mid C, X=x_{0}+\epsilon, Z=1, T=0) \times \mP_{f}(T=0 \mid C, X=x_{0}+\epsilon, Z=1) \\
 & \quad + \mE_{f}(Y \mid C, X=x_{0}+\epsilon, Z=1, T=1) \times \mP_{f}(T=1 \mid C, X=x_{0}+\epsilon, Z=1) \\
 & \quad - \mE_{f}(Y \mid C, X=x_{0}-\epsilon, Z=0, T=0) \times \mP_{f}(T=0 \mid C, X=x_{0}-\epsilon, Z=0) \\
 & \quad - \mE_{f}(Y \mid C, X=x_{0}-\epsilon, Z=0, T=1) \times \mP_{f}(T=1 \mid C, X=x_{0}-\epsilon, Z=0)] \\
 & = \lim_{\epsilon \downarrow 0} \{ E_{f}(Y \mid C, X=x_{0}+\epsilon, Z=1, T=0) \times [1-\mP_{f}(T=1 \mid C, X=x_{0}+\epsilon, Z=1)] \\
 & \quad + \mE_{f}(Y \mid C, X=x_{0}+\epsilon, Z=1, T=1) \times \mP_{f}(T=1 \mid C, X=x_{0}+\epsilon, Z=1) \\
 & \quad - \mE_{f}(Y \mid C, X=x_{0}-\epsilon, Z=0, T=0) \times [1 - \mP_{f}(T=1 \mid C, X=x_{0}-\epsilon, Z=0)] \\
 & \quad - \mE_{f}(Y \mid C, X=x_{0}-\epsilon, Z=0, T=1) \times \mP_{f}(T=1 \mid C, X=x_{0}-\epsilon, Z=0) \} \\
  & = \lim_{\epsilon \downarrow 0} \{ E_{f}(Y \mid C, X=x_{0}+\epsilon, T=0) \times [1-\mP_{f}(T=1 \mid C, X=x_{0}+\epsilon, Z=1)] \\
  & \quad + \mE_{f}(Y \mid C, X=x_{0}+\epsilon, T=1) \times \mP_{f}(T=1 \mid C, X=x_{0}+\epsilon, Z=1) \\
  & \quad - \mE_{f}(Y \mid C, X=x_{0}-\epsilon, T=0) \times [1 - \mP_{f}(T=1 \mid C, X=x_{0}-\epsilon, Z=0)] \\
  & \quad - \mE_{f}(Y \mid C, X=x_{0}-\epsilon, T=1) \times \mP_{f}(T=1 \mid C, X=x_{0}-\epsilon, Z=0) \}\text{~~(from (16))} \\
  & = \lim_{\epsilon \downarrow 0} \{ \mE_{f}(Y \mid C, X=x_{0}+\epsilon, T=1) \\
 & \quad \times [P_{f}(T=1 \mid C, X=x_{0}+\epsilon, Z=1)- \mP_{f}(T=1 \mid C, X=x_{0}-\epsilon, Z=0)] \\
 & \quad - \mE_{f}(Y \mid C, X=x_{0}-\epsilon, T=0) \\
 & \quad [ \mP_{f}(T=1 \mid C, X=x_{0}+\epsilon, Z=1)- \mP_{f}(T=1 \mid C, X=x_{0}-\epsilon, Z=0)] \} \quad \text{(Condition 4.9)} \} \\
 & = \lim_{\epsilon \downarrow 0} \{ [\mE_{f}(Y \mid C, X=x_{0}+\epsilon, T=1) - \mE_{f}(Y \mid C, X=x_{0}-\epsilon, T=0] \\
 & \quad \times [\mP_{f}(T=1 \mid C, X=x_{0}+\epsilon, Z=1)- \mP_{f}(T=1 \mid C, X=x_{0}-\epsilon, Z=0)] \}
\end{align*}
where all the equalities hold $\as{P_{f}}$. 
Rearranging we conclude the proof.
\end{proof}

\subsection{Proof of Theorem 4.14}
\textbf{Theorem 4.14}
\textit{Let $(C,X)$ be strongly sufficient covariates and assume that Condition 4.9 and Property 4.12 hold. Then,}
\begin{equation*}
ACE_{x_0}= \mE_{f}[\lim_{\epsilon \downarrow 0} h(C) \mid X=x_{0}]
\end{equation*}
\textit{where}
\begin{equation*}
h(C)= \frac{\mE_{f}(Y \mid C, X=x_0+\epsilon, Z=1)- \mE_{f}(Y \mid C, X=x_0-\epsilon, Z=0)}{\mP_{f}(T=1 \mid C, X=x_{0}+\epsilon, Z=1)- \mP_{f}(T=1 \mid C, X=x_{0}-\epsilon, Z=0)}. 
\end{equation*}

\begin{proof}
Applying the result of Lemma 4.13 in Theorem 4.11 we conclude the proof.
\end{proof}


\end{appendix}

\end{document}